\documentclass[pdflatex,sn-mathphys-num]{sn-jnl}

\usepackage{graphicx}
\usepackage{color}
\usepackage[figuresright]{rotating}
\usepackage{tabularx}
\usepackage{url}
\usepackage{hyperref}
\usepackage{soul}
\usepackage{booktabs}
\usepackage{textcomp}
\usepackage{verbatim}
\usepackage{amsmath}
\usepackage{amssymb} 
\usepackage{filecontents}
\usepackage{mdframed}
\usepackage{siunitx}
\usepackage{icomma}

\usepackage{subcaption}
\usepackage{caption}

\PassOptionsToPackage{justification=centering}{caption}

\usepackage{adjustbox}

\usepackage{stfloats}

\usepackage{algorithm}
\usepackage{varwidth}
\usepackage{algpseudocode}
\usepackage{xfrac}
\usepackage{enumitem}
\usepackage{comment}
\usepackage[multiple]{footmisc}
\usepackage{tikz}
\usepackage{pgfplots, pgfplotstable}
\usepgfplotslibrary{statistics}

\usetikzlibrary{shapes,arrows,shadows,patterns,positioning,trees,decorations.pathreplacing}
\pgfplotsset{compat=newest}
\usetikzlibrary{arrows.meta}

\algnewcommand{\algorithmicand}{\textbf{ and }}
\algnewcommand{\algorithmicor}{\textbf{ or }}
\algnewcommand{\OR}{\algorithmicor}
\algnewcommand{\AND}{\algorithmicand}
\algnewcommand{\var}{\texttt}

\newlength\myindent
\newcommand{\RN}[1]{%
  \textup{\lowercase\expandafter{\romannumeral#1}}%
}

\tikzset{%
  >=Latex,
  base/.style = {rectangle, rounded corners, draw=black,
                 minimum width=4cm, minimum height=1cm,
                 text centered, font=\sffamily},
  activityStarts/.style = {base, fill=blue!30},
  startstop/.style = {base, fill=red!30},
  activityRuns/.style = {base, fill=green!30},
  process/.style = {base, minimum width=2.5cm, fill=orange!25,
                    font=\ttfamily},
  input/.style = {base, fill=gray!5, text width=4.5cm, font=\ttfamily},
}

\setlist[itemize]{leftmargin=*}

\pgfdeclarelayer{background}
\pgfdeclarelayer{foreground}
\pgfsetlayers{background,main,foreground}

\setlist[itemize]{leftmargin=*}

\theoremstyle{thmstyleone}%
\theoremstyle{thmstyletwo}%

\theoremstyle{thmstylethree}%

\begin{document}


\title[Reproducibility in Event‑Log Research: A Parametrised Generator and Benchmark for Event‑based Signatures]{Reproducibility in Event‑Log Research: A Parametrised Generator and Benchmark for Event‑based Signatures}


 \author[1]{\fnm{Saad} \sur{Khan}}\email{saad.khan@hud.ac.uk}

 \author*[1]{\fnm{Simon} \sur{Parkinson}}\email{simon.parkinson@hud.ac.uk}

 \author[2]{\fnm{Monika} \sur{Roopak}}\email{monika.roopak@beds.ac.uk}

 \affil[1]{\orgdiv{Department of Computer Science}, \orgname{University of Huddersfield}, \orgaddress{\street{Queensgate}, \city{Huddersfield}, \postcode{HD1 3DH}, \state{West Yorkshire}, \country{UK}}}

\affil[2]{\orgdiv{School of Science and Technology}, \orgname{University of Bedfordshire}, \orgaddress{\city{Luton}, \postcode{LU1 3JU}, \country{UK}}}



\abstract{Event-based datasets are crucial for cybersecurity analysis. A key use case is detecting event-based signatures, which represent attacks spanning multiple events and can only be understood once the relevant events are identified and linked. Analysing event datasets is essential for monitoring system security, but their growing volume and frequency create significant scalability and processing difficulties. Researchers rely on these datasets to develop and test techniques for automatically identifying signatures. However, because real datasets are security-sensitive and rarely shared, it becomes difficult to perform meaningful comparative evaluation between different approaches. This work addresses this evaluation limitation by offering a systematic method for generating event logs with known ground truth, enabling reproducible and comparable research. We present a novel parametrised generation technique capable of producing synthetic event datasets that contain event-based signatures for discovery. To demonstrate the capabilities of the technique, we provide a benchmark in signature detection. Our benchmarking demonstrated the suitability of DBSCAN, achieving a score greater than 0.95 Adjusted Rand Index on most generated datasets. This work enhances the ability of researchers to develop and benchmark new cybersecurity techniques, ultimately contributing to more robust and effective cybersecurity measures.}




\maketitle

\section{Introduction}

Event-based signatures are sequences of related system events that are essential for understanding user behaviour, detecting multi-step attacks, and supporting post-incident investigations. Modern IT systems generate large volumes of heterogeneous security events between endpoints, applications, and networks. These events form the basis of security monitoring tools such as Network Security Monitoring systems, Endpoint Detection and Response (EDR) solutions, and Security Information and Event Management (SIEM) platforms, which correlate events to identify complex attack patterns~\cite{navarro2018systematic,levshun2023survey,shaukat2025review}. A key challenge in this domain is that sophisticated threats rarely manifest as isolated atomic events; instead, they produce sequences of related events that collectively form an \textit{event-based signature}~\cite{khan18,bose2013discovering}. 

Detecting such signatures requires linking events that share contextual or object-level relationships, which is a central focus of research on event correlation~\cite{kotenko2022systematic,shaukat2026event} and process mining~\cite{suriadi17}. Object-centric modelling approaches~\cite{ghahfarokhi21} have further improved the ability to reason about these relationships, enabling the extraction of meaningful multi-event structures. However, evaluating techniques for discovering event-based signatures remains difficult in practice. Real-world event datasets rarely include \textit{ground truth} about which events belong to the same activity, and live system data sets are typically security-sensitive and therefore unavailable for open research~\cite{alshaikh22}. Without ground-truth labels or accessible datasets, researchers cannot reliably assess detection accuracy, compare competing techniques, or reproduce experimental results. These challenges continue to surface. The recent exploration of using Large Language Models for event log interpretation has demonstrated the same challenges. It is common for research to be using private data sets that are not shared, creating reproducibility challenges~\cite{bulut2024secencoder}.

Existing datasets and synthetic log generators provide valuable resources, but exhibit several limitations for event-based signature research. Public datasets often focus on anomaly detection rather than signature extraction, and they lack flexibility in terms of completeness, heterogeneity, and structural variation. Synthetic generators used in process mining frequently rely on predefined process models that do not capture the diversity or complexity of security event logs. Consequently, there is no systematic, parametrised method for generating event datasets that (i) contain realistic event-based signatures, (ii) offer controlled variation across key parameters such as signature size, similarity, and frequency, and (iii) provide complete, reproducible ground truth.

This paper addresses these limitations by presenting a parametrised approach for generating synthetic event log datasets designed specifically for event-based signature research. The generator allows users to configure seven parameters that define the structure and scale of the dataset, including the number of signatures, the number of events within each signature, the similarity between events, the number of objects, and the degree of noise. It produces datasets that contain both routine events and structured, labelled event-based signatures. This enables researchers to evaluate signature detection techniques under controlled, repeatable conditions while preserving the characteristics of operational event data.

To demonstrate the utility of the generator, we conducted an extensive benchmarking study using clustering algorithms for event-based signature detection. Preliminary experiments comparing multiple clustering methods identify DBSCAN as the most effective approach in terms of both runtime and accuracy. We then evaluated DBSCAN on more than 12,000 synthetically generated datasets spanning the full parameter space. DBSCAN achieves an Adjusted Rand Index (ARI) above 0.95 on most datasets, and further analysis reveals how specific dataset characteristics influence clustering performance. We also discuss the theoretical limitations of density-based clustering in high-dimensional event–object spaces, linking these insights to the observed failure cases.

The key contributions of this paper are:
\begin{itemize}
    \item A novel parametrised event-log generation technique capable of producing synthetic datasets with configurable event-based signatures and complete ground truth.
    \item A comprehensive benchmarking study involving more than 12,000 datasets, evaluating the performance of clustering-based signature detection and analysing the impact of dataset characteristics on accuracy.
\end{itemize}

This work provides a foundation for reproducible, comparable, and systematic research on event-based signatures and supports the development of more effective cybersecurity analytics.

\section{Background} 
\label{sec:anatomy}

\subsection{What is an Event?} %
\begin{figure}[!t]
\begin{mdframed}
\small
\texttt{
 \begin{tabbing}
Permi\=ssions on an obj\=ect were chang\=ed.\\
\>  ID: log ID 4670\\
Subject:\\
\>	Security ID:	\>	admin\\
\>	Account Name:	\>	John\\
\>	Account Domain:	\>	AD\\
\>	Logon ID:			0x9B3EC\\
Object:\\
\>	Object Server:\>	Security\\
\>	Object Name:\>	D:\\\
\>	Handle ID:	\>	0x5bc\\
Process:\\
\>	Process ID:	\>0x1820\\
Permissions Change:\\
\>	Original Security Descriptor:\>\>	\\D:(A;OICI;FA;;;SY)(A;OICI;FA;;;BA)\\
\>	New Security Descriptor:\>\>	\\D:ARAI(A;OICIID;FA;;;SY)(A;OICIID;FA;;;BA)\\(A;OICIID;FW;;;bob)\\
 \end{tabbing}
}
\end{mdframed}
\caption{Example Microsoft event (ID 4670) detailing the change in security permissions}
\label{fig:microsoftevent}
\end{figure}

\begin{figure}[!t]
\begin{mdframed}
\small
\texttt{
 \begin{tabbing}
Apr 28 17:06:20 ip-172-31-11-241 sshd[1247]:\\ Invalid user admin from 216.19.2.8
 \end{tabbing}
}
\end{mdframed}
\caption{Example Syslog event for an invalid user login}
\label{fig:syslogevent}
\end{figure}

An {\it event} is a discrete record generated by the operating system or running application to record information of relevance for tasks such as future auditing, debugging, etc. Events often contain concise information relevant to the activity and are stored as either key-value pairs or as unstructured text data types. The amount of detail, format, and content of the event differ depending on the vendor/developer and the specification. Figure~\ref{fig:microsoftevent} provides an example event description for assigning new file system permissions on Microsoft operating systems. As in the text details, the event ID is 4670, and the series of objects that were involved or affected by the activity are listed below. This includes those related to ``Subject'', ``Object'', ``Process'', and those related to ``Permissions Change''. An example object is ``Account Name'', and the object value is ``John''. On the other hand, Figure~\ref{fig:syslogevent} presents an example event generated by the Linux system in the Syslog source. The event has been generated to report an invalid user admin login. Unlike the Microsoft format, this event is without type and a key value definition. However, as the format is repeatable due to following a specific template, i.e., the account name always follows the `user' keyword, it can be converted to a key-value representation.

\subsection{Object-Centric Model}
\label{sec:events}
Event logs can be recorded in different formats depending on their source, service, or vendor. To enable efficient processing and analysis, a unified data structure is needed that can store and represent diverse log formats consistently, regardless of their origin. Every event logging system follows a certain template or predefined format; the overall structure of the log entry remains consistent, and the system fills in the placeholders with the relevant values. This consistency in structure makes it easier for systems to parse event logs efficiently. Therefore, it is possible that each event, regardless of its source, can be modelled into the same `object-centric' representation that consists of key-value pairs where the keys represent different attributes of an event (e.g., date, time, username) and the values represent the specific information for each attribute. 

Based on previously published works~\cite{ghahfarokhi21,khan18,khan19}, information in the event logs can be modelled in the {\it object-centric} manner as follows:
\begin{itemize}
    \item $D$ is used to model the set of event entries, where $D = \{E_1 , E_2, ..., E_n\}$.
    \item The event $E = \{T,ID,O\}$, where $T$ is a timestamp, $ID$ represents the type of event in a numeric format, and $O$ is the set of objects such that $O = \{O_1, O_2, ..., O_N\}$.
    \item Each entry, $E_i$, contains objects from $O$ that are involved or affected by the event.   
\end{itemize}

It is important to clarify here that certain event sources do not explicitly format or represent data in key-value pairs; instead, the key-value relationships are inherent there and must be parsed/extracted during analysis. For example, the event entries of a Syslog file include values relating to the keys of users, IP addresses, services, etc. This can be viewed in an object-oriented manner. In addition, the event entries do not contain any information about the event type. Solving these issues would require the implementation of a tailored pre-processing technique for each different event log format~\cite{marin2021event} to translate them into the necessary key-value structure. However, this is only required for some definitions, as most are already in a key-value structure. 

For example, as seen in Figure~\ref{fig:microsoftevent}, in Microsoft events the key-value pairs are easily extracted, whereas in Figure~\ref{fig:syslogevent}, the objects are provided in a string that is more like a natural sentence. In addition, we acknowledge that there are open challenges in how an object model can be parsed without prior knowledge, handling missing data, and heterogeneity~\cite{he2016evaluation}. Several solutions are now available that are capable of learning and extracting objects from the event log without human input~\cite{zhu2019tools,goossens2024object}. 

These topics fall outside the scope of this article; therefore, we are unable to delve into it. The focus of this paper is the generation and analysis of synthetic event datasets, so we assume that the knowledge of modelling structured or unstructured logs into object-centric representation (key-value pairs) is known. 

In this work, it is necessary to have a common event structure that is used to generate and store events. The common structure must be sufficiently flexible to store events extracted from different proprietary formats. Thereby, we continue with the notation of events consisting of a series of objects containing key-value pairs.

\subsection{Event-based Signatures}
\label{evtchainsexp}
\begin{figure*}[!t]
\centering
\includegraphics[scale=0.4]{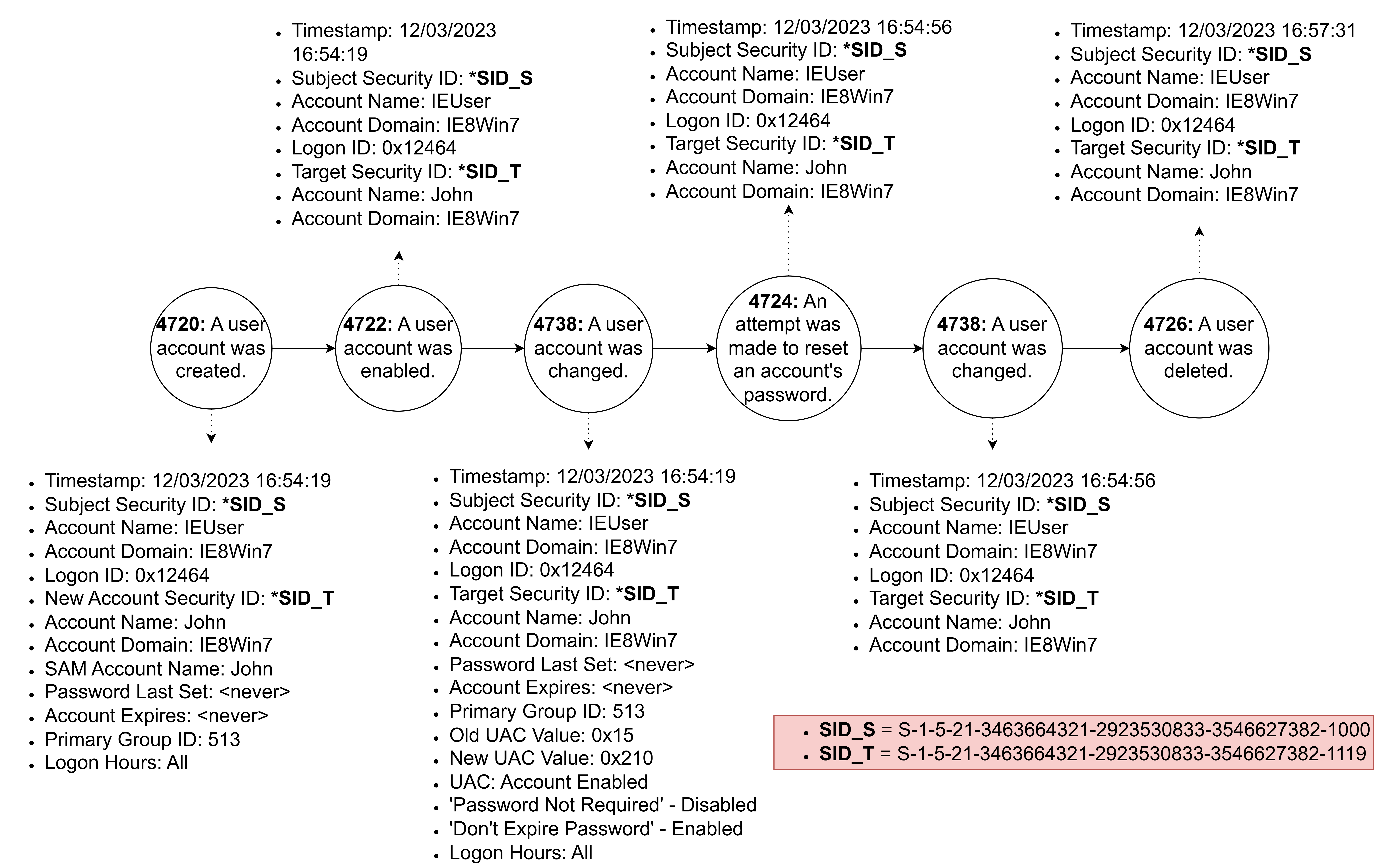}
\caption{Example of a suspicious system activity. The labels SID\_S (Source Security Identifier) and SID\_T (Target Security Identifier), which identify the source and target entities involved in the activity, are provided in the figure's bottom-right corner.}
\label{fig:example}
\end{figure*}

Event-based signatures consist of sequences of events that are grouped together because they collectively represent a specific user, application, or network \textit{activity} or \textit{process}. They denote a temporal sequence of events that are connected in terms of event information included within their contents. These signatures help identify and analyse patterns of behaviour within a system. For example, Figure~\ref{fig:example} presents a real Microsoft event-based signature that involves five types of events (4720, 4722, 4738, 4724, and 4726). This signature represents one of the common approaches to conceal malicious activities, which is to create a user account, launch an attack through the new account, and then delete it immediately to remove all data and traces~\cite{al2017forensic}. 

It is evident that the events in the sequence share several objects. In terms of object-centric representation, although the relationships between these event types may not be encoded, the events share descriptive content. More specifically, there is strong object co-occurrence among events. These objects provide valuable information that can be matched to identify connections between events. For example, user- or application-specific information, such as IDs and names, can be used to determine relationships between events associated with a particular user or an application. It is worth adding here that in previous research related to mining event signatures, researchers often refer to the number of shared objects as a measure of {\it relationship strength}~\cite{khan18,diba2020extraction,goossens2022enhancing}, which helps to discover patterns and anomalies in event log data. The quality and completeness of the objects significantly influence the effectiveness of the determination of the relationship. In the example, there are 17 unique objects throughout the signature, all events share a common set of 7 objects, and only two events have objects beyond the common 7 (4720 and 4738).

\section{Related Work}
\label{sec:related}




Security operations increasingly depend on heterogeneous event data sources (host, application, and network) to detect threats, reconstruct activity, and support investigations. A persistent obstacle for developing unsupervised correlation methods (such as non-parametric hierarchical clustering) is the scarcity of datasets with verifiable ground truth about which events belong to the same user, application, activity, or attack stage. This section reviews (i) security event datasets commonly used to evaluate log analysis methods and (ii) synthetic log generators for security benchmarking, highlighting the limitations that motivate our problem formulation and method. Table~\ref{tab:event_review} summarises key information from publicly accessible datasets discussed in this section.

\begin{table*}[!t]
\centering
\resizebox{\textwidth}{!}
{\begin{tabular}{|l|l|l|} \hline
    \textbf{Research Study} & \textbf{Characteristics} &  \textbf{Size}    \\ \hline           
    Alzhrani et al. 2023~\cite{alzhrani2023bela} & Decentralised blockchain-based application & 101 datasets\\ \hline
    {\v{S}}pa{\v{c}}ek et al. 2022~\cite{vspavcek2022encrypted} & Network and host-based event logs & 21 datasets\\ \hline
    Wu et al. 2022 \cite{wu2022joint} & Network and system, host event logs, RAS log messages & 2,000,000 entries \\ \hline
    Kara et al. 2021~\cite{kara2021read} & Real cyber attack on a Microsoft system  & Not available \\ \hline
    Cinque et al. 2020~\cite{cinque2020contextual} & Air traffic proprietary log data consisting of application and system logs &18 datasets\\ \hline
    De Leoni et al. 2015\cite{de2015road} & Event log of an information system managing road traffic fines & 150,270 events \\ \hline
    Mannhardt 2017~\cite{mannhardt2017hospital} & Event log was obtained from the financial modules of the ERP system of a regional hospital & 177,751 events \\ \hline
    Buijs 2014~\cite{buijs2014receipt} & Event logs of execution of building permit & 1,348 events  \\ \hline
    Steeman 2013~\cite{steeman2013bpi} & Logs of Volvo IT incident and problem management & 1,236 events  \\ \hline
    Mannhardt 2016~\cite{mannhardt2016sepsis} & Event log contains events of sepsis cases from a hospital & 266 events  \\ \hline
    Turcottee et al. 2019 \cite{turcotte2019unified} & Host Event logs & 23,000,000  entries  \\ \hline
    Khan et al. 2018,2019~\cite{khan18, khan19} & 20 end-point event logs  & 3,000-27,000 events  \\ \hline
    Zhu et al. 2023~\cite{zhu2023loghub} & Collected from various system & 509,430,034 entries  \\ \hline
    Flynn \& Olukoya 2025~\cite{flynn2025using} & Real and synthetic network event logs & 186,973 entries  \\ \hline    
\end{tabular}}
\caption{Characteristics of Event log datasets used in the existing and related research}
\label{tab:event_review}
\end{table*}

\subsection{Security Event Datasets for Evaluation}

Several works aggregate multi-source security telemetry to support detection and prediction tasks. Wu et al.~\cite{wu2022joint} evaluated a neural model using four real-world datasets: \emph{Multilog} (system/network/file monitoring with attack-related data), \emph{WAF log} (web-application firewall alerts), \emph{ARCS} (enterprise Windows security events), and \emph{CFDR Blue} (Blue Gene/P RAS logs). These collections provide breadth across host, network, and application layers, but exhibit heterogeneous schemas, partial labelling, and limited cross-source linkage—constraints that complicate the objective assessment of unsupervised event correlation. 

Loghub~\cite{zhu2023loghub} curates various system logs (e.g. Windows, Linux, Mac OS, OpenSH, HDFS, Spark, OpenStack, Blue Gene/L), enabling benchmarking of log parsing and anomaly detection. Follow-up studies used this diversity for security analytics~\cite{mantyla2024loglead,xie2024logsd}. However, many logs on Loghub target operational scenarios and lack explicit ground truth for actor, session, or activity-level grouping needed to evaluate clustering-based correlation.

Domain-specific event sets can expose realistic security signals, yet often lack standardised annotations. For example, encrypted traffic and web hosting telemetry from university IIS servers~\cite{vspavcek2022encrypted} capture rich web activity but offer limited ground truth to attack beyond raw events. Datasets drawn from real cybercrime cases~\cite{kara2021read} contain system, application, and network logs directly relevant to forensics, although access restrictions and non-uniform formats hinder reproducibility. Air-traffic control system events~\cite{cinque2020contextual} include benign operations alongside SSH scans and DoS incidents, but domain specificity may limit generalisation.

A widely used enterprise-scale resource is the LANL unified dataset~\cite{turcotte2019unified}, which integrates host event logs and network flows over ~90 days, with careful de-duplication and anonymisation. It remains valuable for threat detection research; nevertheless, anonymisation and incomplete linkage metadata reduce semantic interpretability and restrict fine-grained ground truth (e.g., precise actor/application/activity associations). Complementary traffic-orientated benchmarks, such as IoT-23 (employed in~\cite{flynn2025using}), provide labelled benign/malicious classes across attack types (DDoS, C2, scans), yet they primarily capture flow/packet behaviours rather than host/application event semantics.

In summary, public security datasets cover important surfaces (host, network, and application), but none provide comprehensive event-level ground truth for grouping events by user/application/activity across sources. This hampers the objective evaluation of unsupervised correlation methods and motivates the use of synthetic data where the causal structure and labels are controlled~\cite{sommers2025groundtruthapproachassessing}.

\subsection{Synthetic Event Log Generation for Security}

To address limited ground truth, synthetic generators are used to benchmark algorithms. General-purpose process mining generators such as PLG2~\cite{burattin2015plg2} and the purpose-guided PURPLE~\cite{burattin2022purpose} enable control-flow design and targeted trace creation; SAMPLE~\cite{gruger2023sample} extends this by adding a data perspective, and SynLogGen~\cite{esgin2019process} supports profile-driven generation using Petri-net abstractions. Subsequent work argues for more modular, extensible generation with richer constraints~\cite{pradhan2025getting}. While influential, these tools are \emph{business-process centric} and under-model security-specific entities (e.g., users, hosts, processes), multi-source causality, and adversarial tactics/techniques. As a result, they do not provide the ground truth of the cross-source activity needed to assess the correlation of unsupervised events in security telemetry. 

Other generators target IoT or generic data synthesis. The IoT Process Log Generator~\cite{zisgen2022generating} produces sensor event logs with configurable noise and ground truth; however, it does not model cyber kill chains, privilege contexts, or host/application interactions typical of security events. MC-GEN~\cite{li2023mc} focusses on privacy-preserving tabular data through multilevel clustering and differentially private Gaussians, and CoSMo~\cite{oyamada2024cosmo} learns control/data-flow patterns for conditioned simulation; both improve realism for process analytics but provide limited control over security semantics (e.g., user-process-resource relations, TTP injection, policy/constraint enforcement).

Synthetic Security-specific efforts have advanced primarily at the \emph{network} layer. Surveys report the utility of GAN variants for realistic attack traffic synthesis and training IDS models~\cite{agrawal2024review}; comparative studies show that GANs can outperform classical and statistical AI methods on fidelity and downstream utility using NSL-KDD and CICIDS-2017~\cite{ammara2024synthetic}; hybrid approaches (e.g., GAN + Random Forest) improve precision in UNSW-NB15~\cite{rahman2025leveraging}. Yet, these benchmarks mainly reflect flow/packet distributions and \emph{do not} capture host/application event semantics or event-to-activity ground truth. Consequently, they are ill-suited to evaluating unsupervised clustering that aims to group heterogeneous \emph{events} into coherent activities or actor-driven threads.

Existing security datasets either lack detailed cross-source ground truth for event-to-activity linkage, or emphasise domains (e.g., network flows) that omit host/application semantics. Meanwhile, general-purpose generators primarily model control-flow, with limited support for security entities, policies, and adversarial behaviour. This leaves a gap for evaluating \emph{unsupervised} correlation methods that must operate without prior knowledge of the number or shape of latent activity clusters. Our work addresses this gap by focussing on non-parametric hierarchical clustering tailored to heterogeneous security logs—explicitly targeting robustness to unknown structure and high dimensionality—and by adopting an evaluation protocol that prioritises datasets and synthetic scenarios where ground truth for activity-level grouping and cross-source linkage can be verified.

\section{Dataset Generation}
\label{sec:datagenerator}

Regardless of the application area, artificially generated data should mimic realistic data; otherwise, algorithms trained on or using the synthetic data may have limited usability once deployed. Synthetic event logs must closely resemble real-world event logs in critical aspects such as timestamp distributions, event relationships, variability, noise levels, and scalability~\cite{levshun2023survey}. While this paper primarily evaluates clustering performance using ARI, future work will incorporate realism validation by comparing statistical characteristics (e.g., object frequency, event lengths) with real-world datasets. Alternatively, fidelity evaluation techniques discussed in existing works such as~\cite{lee19} may be adopted to strengthen realism assessment.

Several techniques are available that can be employed to generate synthetic event log data. Common generator tools are based on: human experts, rules or templates created by observing the patterns in real-world event logs, eliciting and utilising the statistical/parametric properties of real-world event logs, trained machine learning models that can replicate the structure of real-world event logs, data augmentation, etc. Human expertise is in short supply, and the produced synthetic event log generator might not generalise well to other domains or applications. Rule- or template-based systems require manual extraction and encoding of rules, which is a labour-intensive task. Machine learning models might generate inaccurate results if trained on biased, incomplete, or low-quality data. The data augmentation approaches are limited to augmenting existing data and might include biases. Therefore, in this work, we have adopted a parametric approach as it is capable of capturing properties and distributions of real-world event log data. It does not require any input from human experts or existing real-time event logs and can generate entirely new data.

Extensive research has been conducted on the use of event log analysis for various applications. These include knowledge mining~\cite{djenouri2018extracting}, system failure identification and prediction~\cite{arun2024efficient}, anomaly detection~\cite{guo2024logformer}, forensic analysis~\cite{khan2023context} and correlation/causal analysis~\cite{levshun2023survey}. We have selected the following parameters for our synthetic event log generator based on existing studies identified in Section~\ref{sec:related}. These parameters provide sufficient flexibility to accommodate a wide range of scenarios, test cases, and configurations.

\begin{enumerate}
    \item {\bf Events} -- Real-world event logs vary in size depending on system usage, logging policies, and time duration. It is important to generate synthetic event logs of different sizes, small and large, to ensure that they align with the expected scale of real-world systems.
    \item {\bf Objects} -- Each entry in the event log is composed of multiple objects or attributes (e.g., in key-value pairs format). Synthetic event logs should mimic the complexity and diversity of real-world event logs by including a sufficient number of distinct objects.
    \item {\bf Event-object relationship} -- There are different types of events that represent different types of user, system, application, and network activities. Each event type has a different number of objects that provide complete information about the event that occurred. It is important to have the ability to specify the number of objects per event type in the synthetic data to replicate the variability of real-world event logs. This ensures that the synthetic events exhibit realistic details.
    \item {\bf Event signatures} -- Many activities within a system involve multiple steps, where each step will trigger one or more events that are related/connected. In addition, activities often depend on or interact with multiple system components, triggering multiple related events. In addition, when errors occur during an activity, they are logged as separate events to allow troubleshooting. Therefore, it is important to simulate signatures of related events in synthetic event logs, depicting the structure of real-world event logs. 
    \item {\bf Events per signature} -- The number of events varies within related event signatures, representing different scenarios or processes. Therefore, it is important to allow for different numbers of related events in a signature to model activities of different sizes observed in real-world data.
    \item {\bf Similarity within signature events} -- Another important aspect to consider while generating synthetic event logs is to allow adjustable levels of similarity among events within a signature of related events. Some related events have more in common than other related events, for example, due to belonging to the same functional area. This simulates the heterogeneity of related events found in real-world logs, ranging from highly consistent to more diverse relationships.
    \item {\bf Signature repetition} -- Some activities occur more frequently than others in real-world event logs, depending on the system usage. Therefore, it is important to allow for an adjustable frequency of occurrence for each related event signature. This enables the replication of activities ranging from highly frequent to highly rare in the synthetic data, ensuring that it accurately represents the frequency distribution of related event-based signatures observed in real-world scenarios.
\end{enumerate}

In this paper, we present a flexible tool for generating event log datasets. Note that, unlike existing solutions, our proposed solution does not require additional input, such as a process model. The only input required is the specification of seven numeric parameters as seen in the circular shapes in Figure~\ref{fig:process}. The tool generates an event dataset file following the {\it object-centric} approach presented in Section~\ref{sec:events} and contains both noise and event-based signatures (i.e., connected events representing different actions to perform a single task or an activity), which is the same as what is stored in an event dataset produced by a live system. In conjunction with the event log file, our technique provides complete ground truth for benchmarking and evaluation purposes. This is an important and fundamental aspect missing in event datasets publicly shared and those acquired from a live system. This approach to generating an event dataset ensures its suitability for wider practical applications. Figure~\ref{fig:process} illustrates the process and is used to complement the following discussion.

\tikzstyle{materia}=[draw, fill=blue!20, text width=6.0em, text centered,
  minimum height=1.5em,drop shadow]
\tikzstyle{etape} = [materia, text width=8em, minimum width=10em,
  minimum height=3em, rounded corners, drop shadow]
\tikzstyle{texto} = [above, text width=6em, text centered]
\tikzstyle{linepart} = [draw, thick, color=black!50, -latex', dashed]
\tikzstyle{line} = [draw, thick, color=black!50, -latex']
\tikzstyle{ur}=[draw, text centered, minimum height=0.01em]

\newcommand{\blockdist}{1.3}
\newcommand{\edgedist}{1.5}

\newcommand{\etape}[2]{node (p#1) [etape]
  {#2}}

\newcommand{\background}[5]{%
  \begin{pgfonlayer}{background}
    \path (#1.west |- #2.north)+(-0.75,0.75) node (a1) {};
    \path (#3.east |- #4.south)+(+0.75,-0.75) node (a2) {};
    \path[fill=yellow!20,rounded corners, draw=black!50, dashed]
      (a1) rectangle (a2);
      \path (#3.east |- #2.north)+(0,0.25)--(#1.west |- #2.north) node[midway] (#5-n) {};
      \path (#3.east |- #2.south)+(0,-0.35)--(#1.west |- #2.south) node[midway] (#5-s) {};
      \path (#3.east |- #2.north)+(0.7,0)--(#3.east |- #4.south) node[midway] (#5-w) {};
  \end{pgfonlayer}}

\newcommand{\transreceptor}[3]{%
  \path [linepart] (#1.east) -- node [above]
    {\scriptsize #2} (#3);}

\tikzstyle{cloud} = [draw, ellipse,fill=red!20, node distance=3cm,
    minimum height=2em,align=center,text width=2cm]
    
\begin{figure*}
\small
\centering
\begin{tikzpicture}[scale=0.7,transform shape]
  \path \etape{1}{Generate Objects};
  \node [cloud, left of=p1, xshift=-15mm] (n1) {(2) object \#};
  \path[->] (n1) edge (p1);
  \path (p1.south)+(0.0,-2.0) \etape{2}{Generate Events};
  \node [cloud, left of=p2, xshift=-15mm] (n2) {(1) event \#};
  \path[->] (n2) edge (p2);

  \path[every node/.style={}] (p2.north)+(0.0,0.5) node[] {Signature Generation} (p2.north);
  \path (p2.south)+(0.0,-1.0) \etape{3}{Generate Signatures};
  \node [cloud, left of=p3, xshift=-15mm] (n4) {(5) events per signature};
  \path[->] (n4) edge (p3);
  \node [cloud, right of=p3, xshift=+15mm] (n5) {(4) \# of signatures };
  \path[->] (n5) edge (p3);
  \node [cloud, above right of=p3, xshift=+25mm] (n5) {(7) \# of signature repetition};
  \path[->] (n5) edge (p3);
  \path (p3.south)+(0.0,-1.0) \etape{4}{Assign Objects};
    \node [cloud, right of=p4, xshift=+15mm] (n3) {(3) objects per event};
  \path[->] (n3) edge (p4);
  \node [cloud, left of=p4, xshift=-15mm] (n6) {(6) \% similarity};
  \path[->] (n6) edge (p4);

  \path (p4.south)+(0,-2.5) \etape{6}{Generate Events};
  \node [cloud, left of=p6, xshift=-15mm] (n7) {(1) \# of events};
  \path[->] (n7) edge (p6);
   \path[every node/.style={}] (p6.north)+(0.0,0.5)  node[] {Generate Noise} (p6.north);

  \path (p6.south)+(0.0,-1.0) \etape{8}{Assign Objects};
    \node [cloud, left of=p8, xshift=-15mm] (n8) {(2) \# of objects};
  \path[->] (n8) edge (p8);

  \path [line] (p1.south) -- node [above] {} (p2.north);
  \path [line] (p2.south) -- node [above] {} (p3);
  \path [line] (p3.south) -- node [above] {} (p4);

  \background{p2}{p2}{p4}{p4}{bk1}
  \background{p6}{p6}{p8}{p8}{bk2}

 \path [line] (p4.south) -- node [above] {} (p6);
  \path [line] (p6.south) -- node [above] {} (p8);

\end{tikzpicture}
\caption{Synthetic Event Generation Process. The red circles are used to illustrate parameter inputs to the approach, blue boxes are used to illustrate functions within the generation process, yellow boxes illustrate functional groupings, and the arrows represent the flow of information.}
\label{fig:process}
\end{figure*}
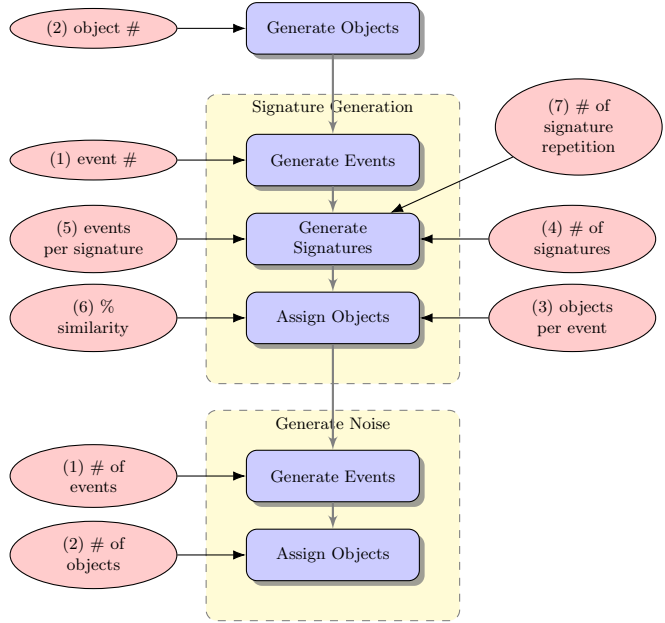

\subsection{Object Generation}
\label{sec:objGen}
In the initial stage, a set of unique objects is generated based on the `number of objects` parameter, each labelled as `objectN`, where `N` is an incremental identifier. These objects are later assigned to events. The process does not generate key-value pairs, as the focus is on developing and testing techniques for identifying event signatures. It is assumed that the raw event log, whether originally structured or unstructured, has already been preprocessed to extract the relevant objects.


\subsection{Event-based Signature Generation}
In the next stage, synthetic event-based signatures are created. These signatures are used to represent the real event-based signatures that will be discovered in the dataset. Our generation process is based on the observation (explained in Section~\ref{evtchainsexp}) that the events in an event-based signature share a relatively high number of common objects compared to events that are not connected. In other words, our tool generates an event-based signature by creating two or more events that have common objects, i.e., have the same object numbers (object1, object2, etc.). Also, note that the number of shared objects in any real-time event-based signature is not predictable and will change depending on the system that generates the events and contextual and situational information. Therefore, it is important to generate datasets with varying amounts of shared objects to capture the realistic event recording mechanism. 


This work structures the generation of event-based signatures into three stages. First, events are created based on a user-defined count. Next, signatures are formed according to the specified number and length, determining how many events each signature contains. Finally, objects are assigned to events based on the required object count and a defined similarity percentage, which controls how many objects are shared across events in a signature. For example, with 10 objects per event and 70\% similarity, 7 objects will be common across all events in the signature. This reflects real-world patterns where events triggered by the same activity share common attributes. Each signature is labelled incrementally (0, 1, 2, ...) to serve as ground truth for evaluating clustering accuracy.

\subsection{Noise Generation}
In addition to the generated signatures, the event dataset also needs to contain {\it routine} or {\it benign} events that represent normal system activity. Such events are not part of any signature related to a security concern and are often recorded as generic and routine descriptive information. In the context of identifying event-based signatures, these events are, in the most part, insignificant. In this work, similar to other works, these routine events are called {\it noise}. 

%
Noise is generated according to the parameters, as demonstrated in Figure~\ref{fig:process}. Firstly, a number of events are generated according to how many events are required to constitute noise in the dataset. The approach generates the required number of events and assigns a fixed number of objects to each. Each of the objects is drawn from those created in the `Generate Objects' activity earlier in the process. As an example, if the technique were to generate 10 events with 10 objects in each, each event would have 10 randomly selected objects assigned. This process ensures that the distribution of objects to events is randomised and best represents noise.

The final event log dataset is created by combining the event-based signatures and noise. It should be noted here that noise and grouped events might have similarities in terms of objects since both types of events are allocated objects from the same set of unique objects (described in Section~\ref{sec:objGen}). Furthermore, depending on the size of the generated event log file, there is a chance that there will be random event patterns in the noise that might be similar to event signatures.

\subsection{Example: From Parameters to Object-Centric Events}
\label{sec:example_object_centric}

To clarify the connection between the proposed generation method and the object-centric model introduced in Section~\ref{sec:events}, we provide a concrete example that maps generation parameters to object-centric event instances. Assume the generator is configured as follows:
\begin{itemize}
    \item Number of event-based signatures: $S=1$ (Signature~1)
    \item Events per signature: $L=3$
    \item Total unique objects: $|O|=8$ with identifiers \texttt{object1}--\texttt{object8}
    \item Objects per event: $m=5$
    \item Similarity within a signature: $70\%$ (i.e., $3$ out of $5$ objects are shared by all events in the signature)
    \item Total number of events: $N_{\text{total}}=9$ (i.e., $3$ signature events and $6$ noise events)
    \item Repetitions per signature: $r=1$
\end{itemize}


We now generate the event signature with $70\%$ similarity and $m=5$, each event in a given signature shares $3$ common objects; the remaining $2$ are event-specific. For illustration, we use plausible keys (\texttt{user}, \texttt{account}, \texttt{host}, \texttt{proc}, \texttt{file}); the generator instantiates object identities from the global pool \texttt{object1}--\texttt{object8}.

\noindent\textbf{Signature 0 (three events, common objects: \texttt{user=object1}, \texttt{account=object2}, \texttt{host=object3})}
\begin{verbatim}
E_01 = { T: 2025-05-01T09:00:00Z, ID: 4720,
         O: { user: object1, account: object2, host: object3,
              proc: object4, file: object5 } }

E_02 = { T: 2025-05-01T09:02:10Z, ID: 4738,
         O: { user: object1, account: object2, host: object3,
              proc: object6, file: object7 } }

E_03 = { T: 2025-05-01T09:04:55Z, ID: 4726,
         O: { user: object1, account: object2, host: object3,
              proc: object4, file: object8 } }
\end{verbatim}

Next noise events are populated with $m=5$ objects sampled from the same pool; they may share objects \emph{by chance} but do not meet the intra-signature similarity constraint:
\begin{verbatim}
E_04 = { T: 2025-05-01T08:50:00Z, ID: 1102,
         O: { user: object7, account: object3, host: object8,
              proc: object5, file: object2 } }

E_05 = { T: 2025-05-01T09:30:00Z, ID: 1201,
         O: { user: object6, account: object8, host: object1,
              proc: object3, file: object4 } }
... (E_06 to E_9 similar)
\end{verbatim}

Each event $E_i$ encodes an \emph{object set} $O$ whose identities are consistent across the dataset (e.g., \texttt{user=object1} appears in all events of Signature~0). Signatures are induced by \emph{object co-occurrence} constraints (here, the $70\%$ similarity rule), not by control-flow alone. This matches directly the object-centric model from Section~\ref{sec:events}: events are linked through shared objects, enabling clustering or correlation to recover signatures. Because both signature events and noise draw from the same $O$ pool, downstream methods must rely on \emph{systematic sharing patterns} (strong intra-signature overlap) rather than incidental overlaps.

In terms of discoverability via shared objects, if we consider a simple similarity function over events that counts shared object identities (e.g., Jaccard over object key--value pairs), then
\[
\text{sim}(E_01,E_02) \approx \frac{\text{shared objects}}{\text{union}} 
= \frac{3}{7} \quad \text{(common: user, account, host)}
\]
and similarly for all pairs inside Signature~0 . These stable, high-overlap relationships create dense neighbourhoods in the feature space, which clustering methods (e.g., DBSCAN) can exploit to recover the ground-truth signatures.

\subsection{Dataset Specification}
\label{lbl:dataspec}

This work generates a range of event datasets using a pragmatic approach to demonstrate the generator's functionality and provide initial benchmarks. Table~\ref{tbl:variables} outlines the input parameter ranges, selected based on prior studies (Section~\ref{sec:related}) and resource constraints. For instance, recent research has processed event data at rates of 5,000 events per hour~\cite{shen18}, and our datasets of 10–40K events represent approximately eight hours of activity. Many studies adopt temporal windowing, such as 1-20 minutes~\cite{lee19}, and limit the number of distinct activities, with one study suggesting a maximum of 14~\cite{fazzinga18}; our work considers 10–40 activities. However, we would like to emphasise that the generator is flexible and capable of producing much larger datasets. In total, over 12,000 synthetic event log files were created to benchmark clustering techniques.

\begin{table}[]
\centering
\begin{tabular}{l|l|l|l}
{\bf Parameter} & {\bf Min} & {\bf Max} & {\bf Stepsize}   \\\hline
Number of event-based signatures & 10 & 40 & 10 \\
Number of events in each signature & 10 & 40 & 10 \\
Number of objects in each event & 5 & 15 & 5 \\
Similarity \% within grouped events of an event signature  & 20 & 80 & 20 \\
Total number of events & 10K & 40K & 10K\\
Total number of unique objects & 100 & 400 & 100 \\
Number of times each signature is repeated & 1 & 4 & 1 \\
\end{tabular}
\caption{Parameter range for benchmark analysis.}
\label{tbl:variables}
\end{table}

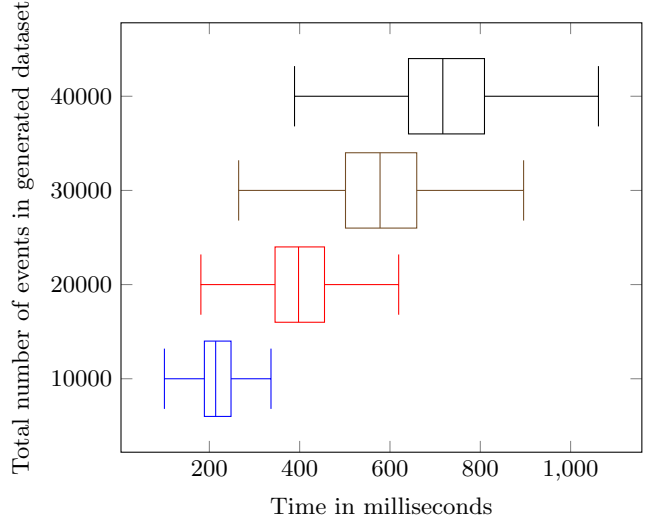
\begin{figure}
\centering
\begin{tikzpicture}
\small
  \begin{axis}
    [
    ytick={1,2,3,4},
    yticklabels={10000,20000,30000,40000},
    xlabel= {Time in milliseconds},
     ylabel= {Total number of events in generated dataset},
    ]
    \addplot+[
    boxplot prepared={
      median=214,
      upper quartile=248,
      lower quartile=189,
      upper whisker=336.5,
      lower whisker=100.5
    },
    ] coordinates {};
    \addplot+[
    boxplot prepared={
      median=397.5,
      upper quartile=455,
      lower quartile=345.5,
      upper whisker=619.25,
      lower whisker=181.25
    },
    ] coordinates {};
    \addplot+[
    boxplot prepared={
     median=578,
      upper quartile=659.25,
      lower quartile=501.5,
      upper whisker=895.875,
      lower whisker=264.875
    },
    ] coordinates {};
    \addplot+[
    boxplot prepared={
      median=717,
      upper quartile=809.25,
      lower quartile=641,
      upper whisker=1061.625,
      lower whisker=388.625
    },
    ] coordinates {};
  \end{axis}
\end{tikzpicture}
\caption{Boxplot showing the time in milliseconds to generate the datasets. The plot focusses on the total number of events in the dataset (10K, 20K, 30K, and 40K).}
\label{fig:gentime}
\end{figure}


Although generating all 12,000 datasets took approximately 20 minutes and 30 seconds, each dataset was created in just 475 milliseconds on average. The process was executed on an Intel i7-8700 CPU @ 3.20GHz with 64GB RAM. Generation time is primarily influenced by the number of events, as larger datasets require more processing. Figure~\ref{fig:gentime} presents a box plot showing generation times for datasets with 10k, 20k, 30k, and 40k events, as listed in Table~\ref{tbl:variables}. This parameter had the most significant impact on generation time, while variation from other parameters is reflected in the box size. For example, datasets with 10k events show a smaller interquartile range than those with 40k. As expected, increasing the number of objects and event signatures also increases generation time. Nonetheless, the results demonstrate that the generation technique is both efficient and scalable.

\section{Clustering Algorithms for Benchmarking}
\label{clusteringalgos}
The following algorithms are selected to evaluate their ability to identify event-based signatures in the datasets generated by the proposed event data generator: Density-Based Spatial Clustering of Applications and Noise (DBSCAN)~\cite{ester1996density}, Ordering Points to Identify the Clustering Structure (OPTICS)~\cite{ankerst1999acm}, Affinity Propagation~\cite{abdulah2022active}, Mean Shift~\cite{zhu2022mean} and Agglomerative~\cite{han2023impact}. We aim to determine how accurate and time-efficient these algorithms are in separating and finding the systematically inserted event-based signatures from noise for benchmarking purposes. The selected clustering algorithms are those that do not require any user input, such as specifying the number of clusters or the size of the neighbourhood. Moreover, we performed several experiments to tune the hyperparameter values to reduce the loss and obtain the best possible performance these algorithms could achieve. The following sections provide information on the algorithm and its configurations determined by preliminary experimentation. These configurations are necessary to replicate our results.

\subsection{Evaluation Metrics} 
\label{section:metrics}

To evaluate the performance of the clustering methods, we have employed the ARI (Adjusted Rand Index), which is an adjusted version of the Rand Index~\cite{hubert1985comparing}. There are other applicable evaluation metrics, such as the Silhouette Score~\cite{shahapure2020cluster} or Normalised Mutual Information~\cite{koopman2017mutual}. The ARI was chosen because it is more appropriate for evaluating clustering performance in event signature detection due to the following key reasons. ARI adjusts for chance grouping of elements, providing a more accurate measure of clustering performance compared to metrics that do not account for random chance. ARI is effective in scenarios with varying cluster sizes and numbers, which is common in event signature detection. ARI allows for direct comparison between different clustering methods on the same dataset, making it easier to evaluate the effectiveness of our proposed method.

RI (Rand Index) is a metric used to measure the similarity between two data clusterings. In the context of this work, the two clusterings being compared are the ground-truth event signatures and those identified using machine learning. Before explaining ARI, it is necessary to first explain how RI is calculated. The RI is calculated as follows:

\begin{center} 
    $RI=\frac{TP+TN}{TP+TN+FP+FN}$
\end{center}

The TP (True Positive) is the number of pairs of event log entries that are in the same subset in the predicted event-based signature and in the same subset in the ground truth (i.e., actual) event signature, which is provided in the labelled input dataset as mentioned before. TN (True Negative) is the number of pairs of event entries that are not in the same subset in the predicted event-based signature and not in the same subset in the ground-truth event signature. FP (False Positive) is the number of pairs of data points that are in the same subset in the predicted event-based signature but not in the same subset in the ground-truth event signature. FN (False Negative) is the number of pairs of data points that are not in the same subset in the predicted event-based signature but in the same subset in the ground-truth event signature. The value of RI ranges from 0 to 1, where 1 means that the predicted and actual event-based signatures match identically.

RI suffers from a major drawback in that it is sensitive to chance~\cite{d2021adjusted}. This means that if the input data is large enough, some data points may be randomly grouped correctly, so the RI might never actually be 0. The Adjusted Rand Index (ARI) considers this possibility~\cite{abdullayeva2022distributed}, so it adjusts/rescales the Rand Index and presents the percentage of correct predictions among all predictions~\cite{sinnott2016chapter}. The Adjusted Rand Index (ARI) is a robust metric for evaluating clustering methods. Unlike unadjusted metrics, ARI accounts for chance groupings, offering a more accurate performance measure. It facilitates comparative analysis by allowing the evaluation of different clustering methods on the same dataset~\cite{krieger1999generalized}. ARI is also effective in handling varying cluster sizes and numbers, which is common in event signature detection scenarios. These attributes justify the choice of ARI despite its less frequent use. ARI is calculated as follows:

\begin{center} 
$ARI=\frac{RI -  Expected_{RI}}{max(RI) - Expected_{RI}}$
\end{center}

Where \textit{Expected\_RI} is the expected value of RI when the clusters (i.e., event-based signatures in our case) are made by coincidence, but while keeping the same marginal distributions, and \textit{max(RI)} is always 1. ARI value ranges from -1 to 1, where 1 indicates a perfect match between the two clusters, 0 indicates a random match, and -1 indicates that the two clusters do not match at all.

Although alternative metrics such as the Silhouette Score~\cite{dudek2019silhouette} or Normalised Mutual Information~\cite{romano2014standardized} are applicable, they are less suitable in this context because Silhouette Score relies on geometric distance separability, which can be misleading in high‑dimensional sparse event‑object spaces, and NMI evaluates similarity between cluster labellings without accounting for chance agreement. In contrast, ARI directly compares predicted clusters against ground‑truth event‑based signatures while adjusting for random matching, making it a more theoretically appropriate metric for this task

\subsection{Algorithm selection}
\label{sec:preliminary}
The event data generator is developed in C\# programming language. The benchmarking experiments are performed using the clustering algorithms implemented in the well-known scikit-learn Python library on a 28-core CPU Intel(R) Xeon(R) Platinum 8180 @ 2.50GHz with 128GB RAM.

To assess the feasibility of running all clustering algorithms in a reasonable time frame, given available computing resources, we conducted a preliminary experiment. Early observations showed that processing every dataset with all algorithms would be too time-consuming. Therefore, we selected a representative subset of 100 event log samples to evaluate the performance of the five algorithms described in Section~\ref{clusteringalgos}. This helped identify which algorithms were suitable for full-scale benchmarking.

\begin{figure*}[t!]
   \centering
    \subcaptionbox{Time in seconds required to cluster the data set. As evident in the Figure, DBSCAN is consistently the fastest.\label{fig:time}}{
        \includegraphics[scale=0.8]{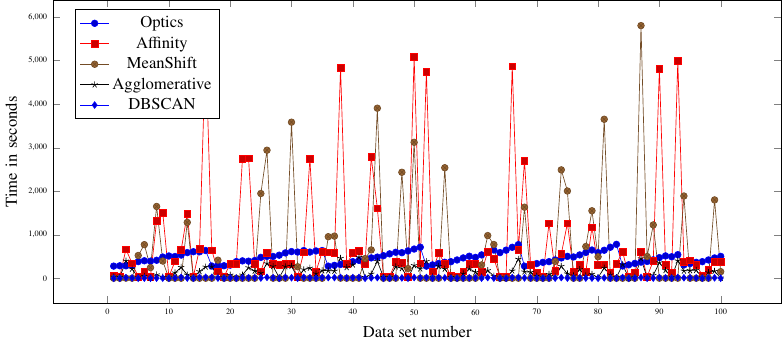}
        }\hfill     
    \subcaptionbox{Clustering algorithm and ground-truth event signature comparison. ARI score of 1 is a perfect match, whereas 0 indicates nothing matching. As seen in the Figure, DBSCAN is the best performing algorithm. \label{fig:rand}}
    { 
        \includegraphics[scale=0.8]{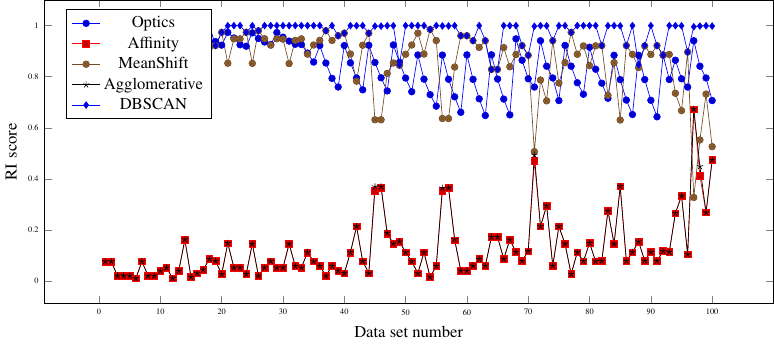}
      }
      \label{fig:time-rand}
    \caption{Performance results of the five selected clustering algorithms using a random selection of 100 datasets from those generated in Section~\ref{lbl:dataspec}}
\end{figure*}

Figures~\ref{fig:time} and~\ref{fig:rand} illustrate the performance of five selected clustering algorithms, evaluated using a randomly chosen set of 100 datasets generated in Section~\ref{lbl:dataspec}.

Figure~\ref{fig:time} compares the execution times (in seconds) across the algorithms. DBSCAN emerges as the fastest, followed by Agglomerative clustering, which despite its speed delivers the poorest ARI scores, as previously discussed. OPTICS ranks third in speed but yields relatively low ARI values, as shown in Figure~\ref{fig:rand}. MeanShift requires significantly more processing time per dataset and also performs poorly in terms of ARI. Affinity clustering is the slowest and produces the lowest ARI scores overall.

Figure~\ref{fig:rand} presents the Adjusted Rand Index (ARI) results for all five algorithms across the 100 datasets. DBSCAN consistently achieves the highest ARI scores, confirming its superior clustering performance. In contrast, Agglomerative clustering performs worst, with ARI values ranging from nearly 0 to a maximum of approximately 0.7. Affinity clustering shows similarly poor results. MeanShift achieves DBSCAN-level ARI scores on only a few datasets, while OPTICS generally performs worse than both MeanShift and DBSCAN.

In summary, DBSCAN demonstrates the best overall performance in terms of both execution time and clustering accuracy. Consequently, it has been selected for benchmark analysis on the generated datasets.

\section{Results and Discussion}
\label{sec:results}
The following section presents the benchmarking results and their detailed analysis of the execution of DBSCAN on all generated datasets. The section is divided into 3 parts. First, the importance of each characteristic is determined to understand the relationship between the characteristics and the adjusted Rand index (ARI) score. This shows which features are relevant and can greatly influence the classification accuracy of DBSCAN. Second, we present and discuss the range of ARI scores for each feature when considering different input values. This provides further insight into the strengths and weaknesses of the algorithm. Finally, we present how multiple features with various input values impact the overall classification accuracy of the algorithm.

\subsection{Determining Feature Importance}


\begin{figure}[!ht]
    \centering
 \includegraphics[scale=0.8]{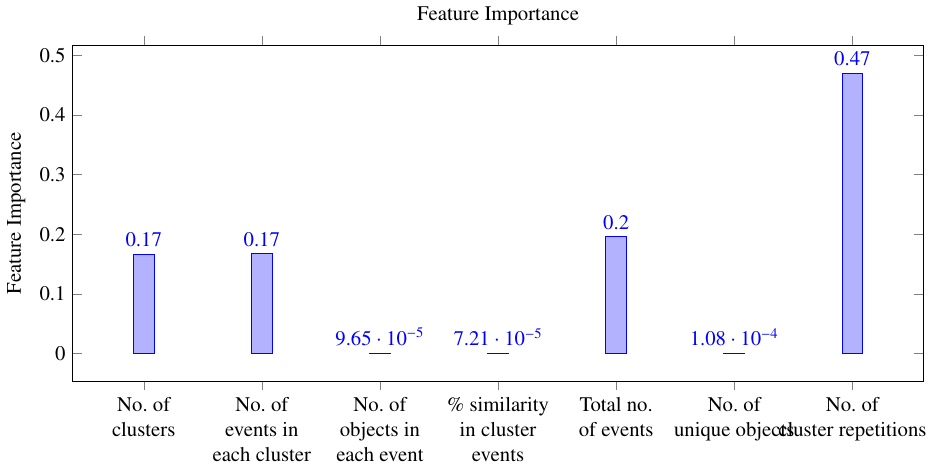}
    \caption{Feature Importance score of all synthetic event generation parameters.}
    \label{fig:importance}
\end{figure}

Based on existing research, we have used the Random Forest (RF) Regressor algorithm to evaluate the importance of each feature present in our generated dataset. Although many feature selection techniques are available, such as those based on linear regression, logistic regression, and neural networks, the RF Regressor has been chosen due to its speed of computation, better accuracy, and effective control over overfitting~\cite{yu2020research,alkuhlani2022glycosylation,saputra2019selection}. As the RF Regressor is based on ensemble learning, it finds the `contribution value' of every feature in every (permuted) decision tree running in the model using the Mean Squared Error (MSE) loss function and then averages all values together, leading to a stronger prediction. A feature is deemed irrelevant if the MSE remains high and unchanged for a given number of iterations~\cite{gromping2015variable}.

Figure~\ref{fig:importance} presents the importance of the characteristic of each attribute corresponding to the target variable (ARI score). The sum of all the numerical values of importance of the feature computed by the RF Regressor is one; the feature with the highest numerical value is considered the most important, the second highest is considered the second most important, etc. It can be seen from the graph that the following three features of our dataset have no significant impact on the ARI score: the total number of unique objects, the percentage similarity in events in the event signature, and the number of objects in each event. The remaining four features are considered significant by the RF Regressor algorithm. The number of times each event-based signature is repeated feature has the highest importance value of 0.47. The number of event-based signatures and the number of events in each event-based signature have a similar influence on the ARI score. The total number of event features is slightly more effective in terms of the numeric value of nearly 0.2 in ARI.

\subsection{Analysis of Each Individual Feature}

\begin{table}[]
\centering
\begin{tabular}{c|c|c|c}
{\bf Variable} & {\bf Min} & {\bf Median}& {\bf Max}\\\hline
\multicolumn{3}{l}{\bf Number of Event Signatures}\\\hline
10& 0.921752& 0.997334& 0.999963\\
20& 0.846705& 0.995121& 0.999928\\
30& 0.774857& 0.993490& 0.999893\\
40& 0.706211& 0.980102& 0.999859\\\hline
\multicolumn{3}{l}{\bf Number of events in each signature}\\\hline
10& 0.921632& 0.995007& 0.999983\\
20& 0.846545& 0.990277& 0.999928\\
30& 0.774737& 0.985062& 0.999841\\
40& 0.706211& 0.980110& 0.999722\\\hline
\multicolumn{3}{l}{\bf Number of objects in each event}\\\hline
5& 0.706211&  0.995007&  0.999921\\
10& 0.706211&  0.995007&  0.999921\\
15& 0.706211&  0.995007&  0.999922\\\hline
\multicolumn{3}{l}{\bf Percentage similarity in each event signature}\\\hline
20& 0.706211& 0.995007& 0.999922\\
40& 0.706211& 0.995007& 0.999921 \\
60& 0.706211& 0.995007& 0.999921 \\
80& 0.706211& 0.995007& 0.999922 \\\hline
\multicolumn{3}{l}{\bf Total number of events}\\\hline
1000& 0.706211& 0.991708& 0.999555\\
2000& 0.846552& 0.996514& 0.999877\\
3000& 0.896245& 0.997785& 0.999944\\
4000& 0.921638& 0.995007& 0.999972\\\hline
\multicolumn{3}{l}{\bf Number of unique objects}\\\hline
100& 0.706211& 0.995007& 0.999921\\
200& 0.706211& 0.994176& 0.999922\\
300& 0.706211& 0.995007& 0.999921\\
400& 0.706211& 0.995007& 0.999921\\\hline
\multicolumn{3}{l}{\bf Number of event-based signature repetitions}\\\hline
1& 0.706211& 0.921638& 0.980103\\
2& 0.998145& 0.999759& 0.999969\\
3& 0.996777& 0.999490& 0.999940\\
4& 0.995442& 0.999176& 0.999875\\
\end{tabular}
\caption{Benchmark results showing the ARI measurement in relation to the changing input variables}
\label{tbl:benchmark}
\end{table}

\begin{figure*}[!h]
\centering
\begin{subfigure}[!ht]{0.4\textwidth}
\centering
\includegraphics[scale=1]{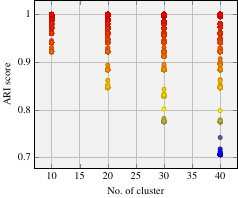}
\caption{The number of event signatures and ARI score}
\label{fig:var1}
\end{subfigure}
\begin{subfigure}[!ht]{0.4\textwidth}
\centering
\includegraphics[scale=1]{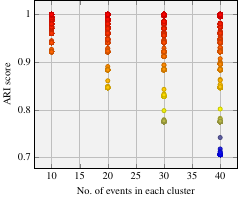}
\caption{The number of events in each signature and ARI score}
\label{fig:var2}
\end{subfigure}
\begin{subfigure}[!ht]{0.4\textwidth}
\centering
\includegraphics[scale=1]{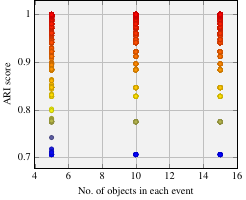}
\caption{The number of objects in each event and ARI score}
\label{fig:var3}
\end{subfigure}
\begin{subfigure}[!ht]{0.4\textwidth}
\centering
\includegraphics[scale=1]{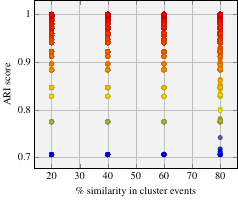}
\caption{Percentage similarity in a signature's events and ARI score}
\label{fig:var4}
\end{subfigure}
\begin{subfigure}[!ht]{0.4\textwidth}
\centering
\includegraphics[scale=1]{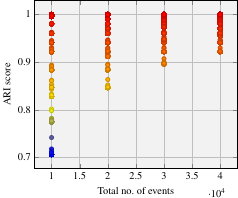}
\caption{The total number of events and ARI score}
\label{fig:var5}
\end{subfigure}
\begin{subfigure}[!ht]{0.4\textwidth}
\centering
\includegraphics[scale=1]{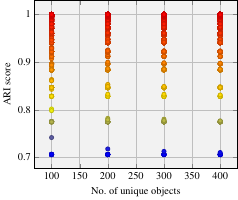}
\caption{The number of unique objects and ARI score}
\label{fig:var6}
\end{subfigure}
\begin{subfigure}[!ht]{0.3\textwidth}
\centering
\includegraphics[scale=1]{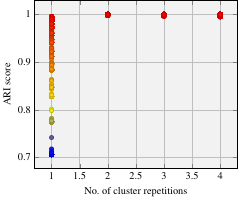}
\caption{Number of times each signatures is repeated and ARI score}
\label{fig:var7}
\end{subfigure}
\caption{Comparison of ARI scores from every individual parameter used for synthetic event generation.}
\label{fig:allvars}
\end{figure*}

The acquired ARI score against each feature is shown in Figure~\ref{fig:allvars} and discussed in the following. As mentioned in Section~\ref{section:metrics}, the ARI has been used as a performance evaluation metric in this work. The value of ARI ranges from -1 to 1, and the higher the value, the more accurate the grouping of data points in the datasets. The value of $ARI \geq 0.95$ is considered optimal~\cite{wang2012constructing}. In addition to the following discussion, the reader is directed to Table~\ref{tbl:benchmark} for the full reference values.

Both the number of event signatures and the number of events in each signature have a similar impact on the ARI score, as is evident from the graphs shown in Figures~\ref{fig:var1} and~\ref{fig:var2}, respectively. Both features have the same numerical values (10, 20, 30, and 40) accordingly and have the same degree of importance. For both features, the value 40 has scored a relatively low ARI, with a minimum of 0.71. The value of 10 has reached the maximum average ARI, with a minimum of 0.91. For values 20 and 30, the minimum scored ARI is 0.85 and 0.78, respectively. These results indicate that increasing the values of these features has a negative impact on the clustering capability of the DBSCAN algorithm. In other words, small event-based signatures with a small number of events result in reduced classification accuracy.

The following 3 features have an almost similar impact on the ARI score: number of objects in each event, percentage similarity in the events of the event signature, and number of total unique objects. This is evident in the graphs presented in Figures~\ref{fig:var3},~\ref{fig:var2} and~\ref{fig:var6} for each of the features, respectively. As previously identified, these characteristics were found to be the least relevant in our generated dataset. The ARI score obtained by these three features ranges from 0.71 to 0.99, and it cannot be specified which feature value(s) always result in poor ARI scores. However, it should be noted that these results, although identified as least significant, still demonstrate that the higher these values, the better the ARI score. This indicates that increasing the number of objects in each event, increasing the similarity within event signatures, and the number of total events all result in the algorithm being more accurate in correctly classifying the event signatures.

Figure~\ref{fig:var5} illustrates the relationship between the total number of unique events and their corresponding ARI scores. The maximum average ARI score is achieved in data files that contain 40,000 events, with a low of 0.91 and a high of 0.99 ARI. For data files containing 20,000 and 30,000 events, their minimum ARI scores are 0.85 and 0.90, respectively. The lowest average ARI score is obtained in the data files where there are only 10,000 events. As identified previously, this is the second most important feature, so increasing the number of events has a positive impact on the ARI scores. This shows that increasing the number of unique objects in the event dataset has a positive impact on the ability to correctly identify event signatures. This is somewhat to be expected, as more unique objects will result in more easily differentiated events.

The ARI produced by changing the number of times each event-based signature is repeated is shown in Figure~\ref{fig:var7}. It is clear that when the value is 1, the average ARI score is the lowest, with a minimum of 0.70. However, when the value is 2 or higher, the average ARI score becomes the highest, which is around 0.99. This feature is of the highest importance in our dataset, so it could be determined that if an event-based signature is repeated multiple times, it becomes easier for the DBSCAN algorithm to accurately identify it. Based on the results and discussion above, it can be concluded that certain individual characteristics influence the accuracy of DBSCAN. However, this might not be a complete picture, as the clustering process is performed using all seven combined features. Therefore, it is important to determine how these characteristics affect the ARI score collectively, which is presented in the next section.


\subsection{Consolidated Analysis of Features}

\begin{table}[!t]
\centering
\begin{tabular}{p{1.5cm}|p{1.5cm}|p{1.5cm}|p{1.5cm}|p{1.5cm}|p{1.5cm}}
 \textbf{Num of event signatures} & \textbf{Num of events in each event signature} &  \textbf{Total number of events} & \textbf{Num of event- based signature repetitions} & \textbf{Average ARI} & \textbf{Range of ARI scores}  \\
\hline 
40  &40 &10000  &1  &0.73     & 0.70-0.74  \\ \hline
40  &30	&10000&	1   &0.77     & 0.77-0.78  \\ \hline
30	&40	&10000	&1	&0.78     & 0.77-0.80 \\ \hline
30	&30 &10000&	1	&0.83     & 0.83-0.84  \\ \hline
40	&20	&10000&	1	&0.87     & 0.85-0.85  \\ \hline
40&	40&	20000&	1&	0.85      & 0.85-0.86 \\ \hline
20	&40	&10000&	1&	0.85      & 0.85-0.86  \\ \hline
40	&40	&20000&	1&	0.86      & 0.85-0.86  \\ \hline
40&	30&	20000&	1&	0.88      & 0.88-0.89  \\ \hline
30	&20	&10000&	1&	0.89      & 0.88-0.90  \\ \hline
30&	40&	20000&	1&	0.89      & 0.88-0.90  \\ \hline
20	&30	&10000&	1&	0.89      & 0.88-0.89  \\ \hline
40	&40	&30000&	1&	0.90      & 0.90-0.91  \\ \hline
30&	30	&20000&	1&	0.91      & 0.91-0.92  \\ \hline
30	&30	&10000	&1&	0.91      & 0.91-0.92  \\ \hline
40&	10&	10000&	1	&0.92     & 0.92-0.92  \\ \hline
40	&20	&20000	&1&	0.92      & 0.92-0.92  \\ \hline
40&	30&	30000&	1&	0.93      & 0.92-0.93  \\ \hline
40	&40	&40000&	1&	0.92      & 0.92-0.92  \\ \hline
30	&40	&30000&	1&	0.93      & 0.92-0.93  \\ \hline
20	& 20&	10000&	1&	0.93  & 0.92-0.93  \\ \hline
20 &	40	&20000&	1	&0.93 & 0.92-0.93  \\ \hline
10&	40	&10000&	1&	0.94      & 0.94-0.94  \\ \hline
40&	30&	40000&	1&	0.94      & 0.94-0.94  \\ \hline
30	&10&	10000&	1&	0.94  & 0.94-0.94  \\ \hline
30&	20	&20000	&1&	0.94      & 0.94-0.94  \\ \hline
30	&30&	30000&	1&	0.94  & 0.94-0.94   \\ \hline
30	&40	&40000&	1&	0.94      & 0.94-0.94  \\ \hline
20&	30	&20000	&1&	0.94      & 0.94-0.94   \\ \hline
10	&30	&10000&	1&	0.94      & 0.94-0.94  \\ \hline
40	&20	&30000&	1	&0.95     & 0.95-0.95   \\ \hline
20	&40	&30000&	1&	0.95      & 0.94-0.95  \\ 
\end{tabular}
\caption{The combination of the four most significant features that resulted in poor ARI scores ($<$ 0.95)}
\label{tab:combination}
\end{table}

This section analyses the four most important features in our dataset -- number of event signatures, number of events per signature, total number of events, and frequency of each event-based signature -- to identify which of their values result in poor ARI scores. This will enable us to gain further insight into the benchmarking results and uncover any general pattern (as to dataset specification) that might lead to poor results. 

The combinations of all feature values, where the ARI scores are less than optimal (i.e., below 0.95), are determined manually by the following process:
\begin{itemize}
    \item List all feature values based on the specification shown in Table~\ref{tbl:variables} and their ARI score obtained by performing DBSCAN on the corresponding data files.
    \item Remove rows from the analysis where the ARI score $\geq$ 0.95.
    \item Sort the list from lowest to highest ARI score.
    \item Group rows together where the four important features have the same values. As each group denotes multiple data files due to having different values of irrelevant features, it contains a range of different ARI scores. 
    \item All such groups and the corresponding range of ARI scores are shown in Table~\ref{tab:combination}, along with their average ARI scores. For example, in a case where the number of event-based signatures is 40, the number of events in each event-based signature is 40, the total number of events is 10,000 and the number of times each event-based signature is repeated is 1, the ARI score ranges from 0.706 to 0.737, averaging at 0.726.
\end{itemize}

Note that the characteristic values not shown in Table~\ref{tab:combination} have achieved ARI of 0.95 or higher, which is considered an optimal score and therefore excluded from the analysis. The reason we are examining the poor results is that they are fewer in number, making them easier to separate and analyse. 

Based on the table, it is evident that the worst ARI score is obtained in those event log files where the number of event-based signatures and the number of events in each event-based signature are the highest (i.e., 40) and the total number of events is the lowest (i.e., 10,000). Therefore, if a number of large event-based signatures that are usually representative of a number of lengthy user, system, network, or application activities are densely packed in a relatively smaller event log dataset, the accuracy is reduced when clustering. This is most likely due to the lack of a clear distinction or separation among the boundaries of event-based signatures (clusters)~\cite{xu2015comprehensive}. If multiple event-based signatures are similar as to event objects and overlap without a drop in density, they are forced into a single event signature, producing poor results.

Another observation is regarding the number of times each event-based signature is repeated, which is always 1 in the case of all poor results. Therefore, if a certain activity is only recorded once within a particular event dataset, it becomes difficult for the clustering algorithm to detect. Finally, there are 1,536 files out of the 12,288 generated that have obtained less than the optimal ARI score (0.95). In other words, only 12.5\% of the generated dataset has obtained an ARI score of less than 0.95. An interesting observation from the results is that the number of files for each combination/group of feature values resulting in a lower-than-optimal score is always 48. Also, note that no log file in the dataset obtained less than 0.7 ARI, which is also considered a good score.

These empirical observations are consistent with the known theoretical limitations of density‑based clustering methods, such as DBSCAN. In high‑dimensional and sparse feature spaces, typical of event‑log representations, the notion of density becomes less meaningful because the distances between points tend to concentrate, making it difficult to distinguish dense regions from surrounding noise. As a result, clusters can merge when there is no pronounced drop in density between neighbouring event‑based signatures, particularly when many large signatures are embedded in a relatively small dataset or when a signature occurs only once. In such cases, DBSCAN may not be able to form clear density valleys between signatures, causing adjacent clusters to be absorbed into a single dense region. This theoretical behaviour aligns with the poor performance observed in datasets with high signature counts, low overall event volumes, and single repetitions.

Considering the benchmark results and the discussion above, it has been established that the application of clustering is suitable for finding and extracting event-based signatures from a noisy event log dataset. Depending on the algorithm, the clustering approach produces accurate results in most cases and is time-efficient. DBSCAN has shown that it can efficiently produce good results.

\section{Conclusion}
\label{sec:conclusion}
This research focusses on the development and description of a tool capable of generating event datasets for the development and analysis of security techniques that process event datasets. 
In this work, we present a parametrised tool capable of generating datasets for a wide range of research objectives centred on the use of event-based signatures. The technique we have produced is defined by 7 parameters to define characteristics of the generated event log, including information about its size, objects, and relationships. In addition, parameters are used to define the characteristics of the event signature. In addition to generating synthetic datasets, the technique can also use real event datasets as a base for adding synthetic signatures with ground-truth knowledge. This combination ensures that the generator is flexible and useful for wide-ranging application scenarios.

We then perform benchmarking on a range of datasets generated by considering different parameter combinations within a predefined range. This paper focusses on the use of clustering to identify event signatures. The initial analysis demonstrated that DBSCAN is the most appropriate algorithm to use in benchmarking, as it can handle the generated datasets in terms of size and processing time. DBSCAN is then applied to all generated datasets, generating promising results and highlighting that clustering provides a suitable approach to event-based signature detection. This paper and the event generation application provide the foundation for future research in this discipline, with a flexible generation technique and datasets that can be shared.

The generation technique is not without limitations and areas for future work. An area to consider is the heterogeneity in the generated datasets. Allowing event sequences to be generated with a similarity range would allow sequences to have variance and to be more realistic to the real world. The reason we decided not to design our approach in this way was because of repeatability. The process of creating events based on a similarity range would likely be generated using a random generator. Although this would indeed help create more interesting event datasets, a dataset might be generated differently multiple times with the same parameters. This would be detrimental to our aim of creating datasets that allow rigorous benchmarking. There is also a great opportunity for researchers to benchmark their approaches on datasets generated using this technique, thus improving the ability to make direct comparisons between competing approaches.


\section*{Declarations}
\begin{itemize}
\item Availability of data and materials: All experimental datasets, scripts and software are available on \url{https://github.com/sparkins01/EventsToActions}
\item Funding: Research funded by Frazer-Nash Consultancy Ltd. on behalf of the Defence Science and Technology Laboratory (Dstl) which is an executive agency of the UK Ministry of Defence providing world-class expertise and delivering cutting-edge science and technology for the benefit of the nation and allies. The research supports the Autonomous Resilient Cyber Defence (ARCD) project within the Dstl Cyber Defence Enhancement programme.
\item Acknowledgements: Not applicable.
\item Competing interests: Not applicable.
\item Authors' contributions: Saad Khan, Simon Parkinson, and Monika Roopak undertook the primary research and drafted the article. All authors contributed equally to this work.
\end{itemize}

\bibliography{bibliography}

\end{document}